# StoryGrid: A Tangible Interface for Student Expression

Tom Moher[1], Louis Gomez[2], Janet Kim[1], Claudia Hindo[2], Benjamin Watson

[1]University of Illinois Chicago

**ABSTRACT**

StorySpace is a classroom-based design and presentation system for interactive multimedia posters. Employing the technology base first used in Eden's PITAboard [2002], StorySpace allows groups of learners to manipulate projected multimedia objects on a horizontal board using a small collection of shared physical tokens. In this paper, we present the ongoing design history of StorySpace in the context of its introduction within an urban high school literature class. Interface modifications based on student and teacher feedback led on changes in token semantics and media importing methods. We describe how StorySpace features enriched students' interpretations of literature, with particular emphasis in two areas: (1) attention to audience, and (2) reflection of multiple perspectives.

**ACM Classification Keywords**

: H.5.1 Multimedia Information Systems; H.5.2 User Interfaces; H.5.3 Group and Organization Interfaces

**INTRODUCTION**

Less than half of adults in the United States read literature: novels, short stories, and poetry. While educators continue to place a premium on the active interpretation of narrative [Wiggins & McTighe, 1998], they appear to be fighting a losing battle: the rate of decline in reading is at an historical maximum and is increasing. In certain subpopulations, the situation is particularly dire; in a survey conducted in 2002, for example, fewer than one in five U.S. Latino males age 18-24 indicated that they read literature [NEA, 2004].

While printed text is not perceived as a high-value medium among students, electronic artifacts have gained ever-greater currency among youth. In recent years, as tools for creating such artifacts have become more accessible, teachers have begun to use audio and video as expressive media in the interpretation of literature (e.g., [hidden for reviewHindo 2004]).

**Figure 1. Students using StorySpace to construct interpretation of Shakespeare's *Macbeth*.**

StorySpace (Figure 1) is a classroom-based design and presentation system for students' interpretation of narrative. StorySpace differs from other student multimedia efforts both in the nature of the resultant work products, and in the way that those products are produced. Instead of single, linearized audio or video pieces, StorySpace adopts a composite framework, in which individual media elements (from commercial audio and video sources, the Internet, and student-captured media) are combined to create interactive posters. The "consumers" of StorySpace products (peers, teachers, parents, etc.) retain the freedom to activate poster elements as they see fit using the same interface as their designers, blurring the distinction between author and audience. By using a tangible, semi-public, multi-user interface, StorySpace supports collaboration and encourages formative critique of student work products.

In this paper, we briefly describe the design history of StorySpace within the context of its usage in an urban high school literature class. Feedback from teachers and students informed design evolution, and shed preliminary light on how the affordances of StorySpace could deepen students' engagement with literature.

**TECHNOLOGY OVERVIEW**

StorySpace employs the same technology base introduced in Eden's PITABoard [2002]. An overhead video projector displays the output of a laptop computer on a horizontal working surface. Underlying that surface is a plastic sheet containing an RFID (radio frequency identification) antenna grid connected to the laptop. Users place custom-built tokens with distinctive "caps" containing RFID *tags* on the working surface; their location and identity is sensed by the antenna and transmitted to the laptop application. Spatial sensing resolution is limited to 8x8 (the antenna grid was designed for a DGT electronic chessboard). A pair of inexpensive speakers were placed in proximity to the board.

The StorySpace software divides the image space into an 8x8 grid of square *cells*, matching the antenna grid, and projects a collection of rectangular multimedia objects onto this grid. Each object is aligned with cell boundaries, and may be scaled in size from a single cell up to the entire board surface. Multimedia objects may overlap one another; the more recently positioned or scaled object is automatically brought to the front, occluding other objects. The tokens are used to change the organization of objects on the board, and to invoke audio and video activity.

QuickTime™ and a
TIFF (Uncompressed) decompressor
are needed to see this picture.

**Figure 2. StorySpace tokens resting on RFID antenna cells.**

Each StorySpace object contains an image component that represents the object in the board's "steady state." While the simplest objects have only this image component, interactive objects have at least one, and sometimes two, additional audio or video components. A configuration application running on the laptop allows users to bind the multimedia components together as a single object, and to specify an initial layout of objects on the board.

**RELATED WORK**

Tabletop tangible interfaces like that used in StorySpace have a surprisingly long history. In the early 1980s, architect John Frazer was frustrated with the inapproachability of the design process and developed the Segal Model [Frazer, 1982], a digital device that allowed architectural clients to prototype their own designs. Users arranged physical panels corresponding to building elements, and output the results on blueprints. Over a decade later, Fitzmaurice, Ishii and Buxton began revisiting these ideas. Their first system, Bricks [Fitzmaurice et al., 1995], allowed users to shape curves and define simple drawings by manipulating physical tokens corresponding to drawing actions. Since then Ishii and his colleagues have created several tabletop interfaces [Ullmer & Ishii, 2001] addressing educational applications in optical design, urban planning and chemistry. Rogers & Lindley [2004] recently found that such horizontal, tabletop surfaces offer advantages over vertical surfaces for collaborative design.

Two systems that have a particularly strong relationship to this project are PITAboard [2002], which was the first to use the DGT chessboard components as a token-tracking interface; and Tangible Viewpoints [Mazalek et al., 2002], which used a tabletop tangible interface to explore multiple character perspectives within narrative.

**DESIGN HISTORY**

**Physical design**

Portability was a central design goal for the system, precluding the use of a ceiling-mounted projector. We used a simple direct projection configuration with the projector mounted on a commercial light standard using a custom-built cantilever. The standard's legs fold in and the cantilever is removable, allowing the system to be collapsed to a small size for storage. Since a long cantilever would have required a much longer base, the standard had to be positioned in close proximity to one of the four sides of the board, reducing the number of usable positions. We also constructed a custom board to hold the RFID antenna and associated circuitry and connectors. White butcher paper was taped on top of the board for the projection surface.

In use, the system proved portable, but setup and breakdown time remain problematic for classroom environments, requiring at least 10-15 minutes each. Since the system was being used only for some class periods and not others, it had to be assembled and disassembled daily, seriously impacting usable time.

Most of the imagery that students captured or developed had a clearly preferred viewing orientation. (In this way, StorySpace differed from the reported PITABoard applications, which typically adopt an overhead view that is orientation-neutral.) As a result, group usage rarely involved the canonical "one user on each side," but instead tended to cluster around the edge reflecting the "proper" orientation. (Ironically, the elimination of the fourth edge as a user position due to the projector mount mitigated the orientation problem.) In future versions of StorySpace, we intend to introduce a *Rotate* token, although it remains an open issue whether rotation should apply to the entire board or to individual media objects.

**Application software**

The StorySpace software was developed first by creating an isomorphic screen-based version of the application so that all of the functionality of the system was available for testing in the absence of the hardware; the same application later became the basis of the board configuration interface. The software interface to the RFID antenna required the development of a custom device driver, which, like the application, was written in Java. The application runs in the Windows XP environment, and utilizes the Windows Media Framework.

A tension between usability and pedagogical goals became evident during usage of StorySpace. Since the configuration tools afforded students full control over object placement and scaling, students tended to do much of their design on the laptop rather than on the board, situating their work around the screen and frustrating the goal of engendering collaborative participation in design. To address this problem, the latest version of the laptop configuration tool may be used to assemble objects from their component parts, but no longer to specify layout. Objects can be arranged and sized only on the board. (Intermediate layouts can be saved and restored, but not manipulated on the laptop.)

**Tokens**

Much of the initial design discussion revolved around the syntax and semantics of the control tokens. In chess, there are six unique playing pieces of each color, requiring only a dozen different tags. Rather than using all of the tags at the outset, we adopted the strategy of initially defining a minimal set of tokens (*Move*, *Grow*, *Shrink*, *Play*, and *Remove*) and employing user feedback to guide revision of the token set. That strategy resulted in significant transformations of the token set over a six-month period.

The basic operations we wished to support were moving, scaling, playing, and deleting media objects. Operations like playing and deleting have a single operand: the media object itself. Thus, a simple syntax—placing an operator on the media object—is sufficient to fully specify the operation. The move operation necessitated a more complex syntax, requiring a pair of gestures to indicate the initial and target locations of a media object. (The technology does not support the equivalent of a "drag" gesture).

We chose a pair of simple, single-operand scaling operators: one to uniformly increase the size of the media object (by one cell in each dimension), and one to uniformly decrease its size.

Pilot tests led to the identification of several usability problems, leading to a number of design revisions. Moving multi-cell objects was confusing for users because of ambiguity in the "anchor" point of translation. We constrained the source and destination points to occupy the same relative positions of the object's image.

The limitations of scaling proved difficult for students, forcing an early decision on object aspect ratio during board configuration. We considered two solutions:

a) Use four tokens for resizing: Grow horizontal, Grow vertical, Shrink horizontal, and Shrink vertical, or

b) Use only one *Resizer* token, and interpret initial selections along media object borders as signaling scaling in the respective dimension and direction.

We chose the second solution. Although it has a complex multi-operand syntax, the solution uses a smaller operator set and allows the user to designate the *direction* of scaling. Other usability driven modifications included the addition of a *Stopper* token to interrupt playback of audio and video files and the introduction of an *Undo* token (for *Move*, *Resize*, and *Delete* operations).

Classroom teachers suggested two additional design modifications. First teachers noted that students frequently resized video objects to the full size of the board, played associated video, and then restoring objects to their original size. This usage pattern was captured in a new token, the *Zoomer*. Applying the Zoomer to an object scaled it to the full size of the board; the next application to that object restored its original state.

Second, teachers suggested that two interpretations of a media object might be supported by linking two distinct files to it. These were originally couched as "student" and "teacher" perspectives, but could be any two. For example, an image of Juliet's paralyzed body could serve as a trigger for both Romeo's tortured soliloquy and Juliet's hopeful anticipation of recovery from the temporary effects of her poison. To effect this we replaced the single *Player* with two separate players labeled simply *Player 1* and *2*.

In group use, students tended to share tokens rather than specializing their roles around retention of an individual token. For one extended design session involving three students, we instrumented the application to create a log of token usage. Design activity was concentrated; the students performed 314 token operations in 82 minutes; excluding two 10-minute breaks, a token operation was performed every 11.8 seconds. The distribution of token usage by type is given in Tables 1 and 2. All of the tokens were used during design, with translation and scaling the most common. The nearly identical usage of the two Player tokens provides evidence of students' attention to multiple perspectives.

**Tables 1 and 2. Frequency of StorySpace token use by a three-student over 82 minutes.**

| Token | Relative Usage Frequency | Token | Relative Usage Frequency |
|---|---|---|---|
| Eraser | 6% | Resizer | 19% |
| Mover | 32% | Stopper | 6% |
| Player 1 | 12% | Undoer | 2% |
| Player 2 | 12% | Zoomer | 11% |

**Student perception of StorySpace**

Anecdotal evidence supports students' affinity for the StorySpace interface. Students comments centered on the tangible interface ("I liked it all, especially like moving the stuff around, using the pieces and how it actually worked the playing it and moving around"), the perceived directness of interaction relative to pointer-mediated input ("with [StorySpace] it's like hands on, so it's like you're basically doing the same thing a computer does except you're doing it more direct yourself. Your focus is more on it than just sitting there and clicking the mouse"), and support for collaboration ("We… took turns on… where we're going to do things… It was better because we could switch off and everything; it would be more easier").

Other aspects of StorySpace were less well received. Aligning the projected image with the RFID antenna required regular projector adjustments. The antenna's sensor grid has "dead spots" between cells, which were problematic for infrequent users; Eden reported similar reactions [2002]. Inconsistencies between the Macintosh capture and editing platform and StorySpace's Windows platform frequently crashed the system in early use.

**EFFECTIVENESS FOR LEARNING**

Motivating engagement with course content can be a significant challenge for teachers. Student comments provided tentative support for StorySpace's value in that effort. One student studying MacBeth told us "If it wasn't for the project I wouldn't know anything about Macbeth, because when I first read the play, I didn't really get it." Another candidly admitted, "I learned a lot from Macbeth and I didn't even read the story."

**Figure 3. StorySpace poster reflecting students' interpretation of Shakespeare's *Macbeth*. The poster contains 23 distinct multimedia objects.**

StorySpace effectively engaged students with their peer audience and encouraged them to accommodate multiple perspectives. Students knew that their StorySpace poster would be interacted with by fellow students, and with this awareness demonstrated an increased sensitivity to both tone and mood. One student reworked the music in his movie because he wanted to convey the "feeling" of poem more effectively to his peers. Students reciting or commenting on the poem on video often made multiple takes in an effort to convey a consistent point of view. StorySpace's interactive tabletop interface allowed students to observe the effect of changes in their poster on their peers, discuss those effects in context with their design partners, and incorporate those collaborative discussions into poster revisions that conveyed multiple perspectives.

## IMPLICATIONS FOR TANGIBLE USER INTERFACES

Though still preliminary, this research shows the promise of tangible user interfaces in education. Students found StorySpace both exciting and accessible. Its tabletop, interactive format encouraged students to collaborate and consider their peers when forming their work, both as an audience and as partners with different perspectives. Yet our experience also revealed some shortcomings of the technology. Tangible user interfaces must become much more robust before they will see widespread use in the classroom. And in a more fundamental challenge, collaborative use was often harmed by the viewing limitations introduced by the preferred orientation in most everyday media. A solution to this problem would be quite valuable in the educational setting.

## ACKNOWLEDGMENTS

We gratefully acknowledge the support and participation of the staff and students at Roberto Clemente High School and the Electronic Visualization Laboratory at the University of Illinois at Chicago. This material is based upon work supported by the National Science Foundation under Grant No. IIS-0112937.